\documentclass[onecolumn,showpacs,showkeys,preprintnumbers,amsmath,amssymb]{revtex4}
\usepackage{setspace}
\usepackage{graphicx}
\usepackage{grffile}
\usepackage{epstopdf}
\usepackage{dcolumn}
\usepackage{bm}
\usepackage{epsfig}

\begin{document}


\preprint{Report IST/WIU 2012-Pinheiro Buker}

\title[]{An Extended Dynamical Equation of Motion, Phase Dependency and Inertial Backreaction}

\author{Mario J. Pinheiro}
\address{Department of Physics, Instituto Superior Tecnico, Av. Rovisco Pais, \& 1049-001
Lisboa, Portugal \& \\
Institute for Advanced Studies in the Space, Propulsion and Energy Sciences,
265 Ita Ann Ln. \\
Madison, AL 35757
USA
} \email{mpinheiro@ist.utl.pt}

\author{Marcus B\"{u}ker}

\affiliation{Department of Geography\\
Western Illinois University, \&
Macomb, IL, USA 61455
} \email{ML-Buker@wiu.edu}




\date{\today}
\begin{abstract}
Newton's second law has limited scope of application when transient phenomena are present. We consider a modification of Newton's second law in order to take into account a sudden change (surge) of angular momentum or linear momentum. We hypothesize that space itself resists such surges according to a kind of induction law (related to inertia); additionally, we provide further evidence of the ``fluidic" nature of space itself. This ``back-reaction" is quantified by the tendency of angular momentum flux threading across a surface. This quantity is mass-dependent, and bears similarity to the quantum mechanics phase shift, present in the Aharonov-Bohm and Aharonov-Casher effects. Furthermore, this provides evidence of vacuum polarization, a phenomena which is relative to local space indicating that local geometry and topology should be taken into account in any fundamental physical theory.
\end{abstract}
\maketitle
\section{Introduction}

Newton's second law has limited scope of application when transient (i.e., jerk or surge) phenomena present. This law states that ``the force is proportional to the rate of change of momentum". In a sense, this implies a kind of ``straight-line" theory, since force and change of momentum are directed along the same line. Critics of the conceptual foundation of this basic equation of dynamics include Gustav Kirchhoff, Ernst Mach, Henri Poincar\'{e}, Max Jammer and Frank Wilczek (see Ref.~\cite{Pinheiro_2009} and Ref.~\cite{Hacyan_09} and references therein).

Tait and Watson~\cite{Tait} delimits the applicability of the concept embedded into the Newton's second law of motion, ``If the motion be that of a {\it material particle}, however, there can be no abrupt change of velocity, nor of direction unless where the velocity is zero, since [...] such would imply the action of an {\it infinite} force".

According to astronomical observations the amount of mass that can be identified in the universe accounts for only a small percentage of the gravitational force that acts on galaxies and other galactic systems. Galaxies' rotation curves show that they do not spin according to Newtonian dynamics; they spin considerably faster. While the standard cosmological model argues dark matter may explain the discrepancies, modification of Newtonian dynamics has been proposed~\cite{Milgrom_83a,Milgrom_83b} as a means of explaining the discrepancy.

The main assumption of the modified-inertia Newtonian dynamics theory (MOND) is that Newton's second law should be modified to read
$\mathbf{F}=m \mathbf{a} \mu(a/a_0)$, where $a_0$ is a fundamental acceleration of the order of $10^{-10}$ m s$^{-2}$ (with respect to the galactic reference frame) and $\mu$ is an unspecified function satisfying the two extreme limits: $\mu (a/a_0) \to 1$, when $a/a_0 \gg 1$, and $\mu (a/a_0) \to a/a_0$, when $a/a_0 \to 0$.

Ignatiev~\cite{Ignatiev_08}, using a version of MOND, proposed a terrestrial test of high-latitude equinox effect hypothesis, which predicts that around each equinox, two spots emerge on Earth causing the spontaneous displacement of static bodies (in clear violation with Newton's second law).

Ref.~\cite{Hacyan_09} gives a general overview of the insufficiency of the standard fundamental equation of dynamics to describe nature, but does not use the approach outlined in this paper. Here, we will show the necessity, at a fundamental level, to modify the usual equation of dynamics in order to describe general mechanical and electromagnetic phenomena.

\section{The 4th Law of Motion}

The concept of self-reaction has been hypothesized in field theory, as early as the classical Abraham-Lorentz, and later the Abraham-Lorentz-Dirac equation (see, e.g., Ref.~\cite{Yaghjian} and references therein).

In this context, considerable progress has been attained in the study of the electric charge dynamics~\cite{Harpaz_1998,Harpaz_2003,Pinheiro}, where the point charge and the extended charge models of the classical electron have been analyzed.

The Abraham-Lorentz non-relativistic equation of motion for a point charge is given by~\cite{Rohrlich,Rohrlich_1}:
\begin{equation}\label{eqa}
m\mathbf{a} = \mathbf{F}^{ext} + m \tau \mathbf{\dot{a}},
\end{equation}
where $\tau=2e^2/(3mc^3)$, $\mathbf{F}(t)=e(\mathbf{E}+\mathbf{v} \times \mathbf{B})$ is the Lorentz force and a dot means a derivative relative to time. Dirac later obtained the equation of motion in a four-dimensional formalism:
\begin{equation}\label{eqa1}
ma^{\mu}=F_{ext}^{\mu}+\Gamma^{\mu}=F_{ext}^{\mu}+\frac{2e^2}{3c^3} \left( \dot{a}^{\mu} - \frac{1}{c^2} v^{\mu} a_{\alpha} a^{\alpha}  \right).
\end{equation}
Here, $\Gamma$ is known as the Schott term and its first term includes the time derivative of the acceleration, already present in the Abraham-Lorentz equation, Eq.~\ref{eqa}.

The last term of the right-hand side (rhs) is the Abraham-Lorentz force term, describing the radiation recoil force due to the self-interaction between the charge and its own force fields. This is an important example of a transient phenomena that appears as an intermediate state between two permanent conditions: in the above instance, the transient is the result of the change of the amount of stored energy in the system (e.g., the extended charge in the above example). This phenomena represents its readjustment to the change of stored energy.

In electrical engineering, when reactive elements are present (for example, in an electric circuit comprising of a solenoid with a given resistance $R$ and inductance $L$), Kirchoff's law dictates the form for the voltage at the terminals:
\begin{equation}\label{eq0}
    V_L=Ri + L \frac{di}{dt}.
\end{equation}
The change in the stored magnetic energy is thus described by the last term involving inductance and the time rate of change of current. Analogously, we can write:
\begin{equation}\label{eq1}
\sum_i \mathbf{F}_i = m\mathbf{a} + \aleph \frac{d\mathbf{a}}{dt},
\end{equation}
where we call $\aleph$ the {\it intractance} (in Eq.~\ref{eq1} in units m$\cdot$s); the other terms have the usual meaning. In the foregoing treatment it will be shown that the last term of Eq.~\ref{eq1} is the time derivative of the ``flux" of angular momentum $\Phi$. This result will verify the assertion of Bjerknes~\cite{Bjerknes} that we need {\it fields} (e.g., $\mathbf{v}$, $\mathbf{L}$) and {\it flux} (e.g., $\mathbf{\Phi}$) to describe fundamental phenomena.

To illustrate the resistance (back-reaction) of space against a sudden change of motion, we will start with an algebraic description of the gyroscopic effect.

The equation of motion for a gyroscope can be obtained by a vectorial method~\cite{Brand_1930,Armstrong_1967,Laithwaite_1,Laithwaite_2,Laithwaite_3}. Consider a wheel spinning around its own instantaneous axis of rotation (and coincident with axis $O_z$) which is free to move about the fixed point $O$ an axisymmetric gyroscope. Let us suppose that the axis of spin is displaced by an infinitesimal angle $d \phi$ in a time $dt$ around axis $O_y$, such as $dL=Ld \phi$. The equation for torque (see also Fig.~\ref{Fig1}) is:
\begin{equation}\label{eq2}
\mathbf{\tau} = \frac{d \mathbf{L}}{dt} = L \mathbf{\Omega} =(I \omega) \mathbf{\Omega}.
\end{equation}
We propose the last term of Eq.~\ref{eq1} is equivalent to a back-reaction force resulting from a mechanical induction law. The implication of Fig.~\ref{Fig1} is that a sudden torque gives rise to the angular momentum represented by a new vector $d \mathbf{L}$ (pointing toward the sheet). However, this sudden action (``surge" or ``jerk"), $d \mathbf{a}/dt$, through an induction mechanism, generates a counter-opposed angular momentum, designated $\mathbf{L}_{ind}$, that appears pointing outside the sheet. This counter-angular momentum is related to the induced torque through the equation (here, its x-component):
\begin{equation}\label{eq3}
\tau_x = (I_z \omega_z)\Omega_y + \tau_{x,ind}.
\end{equation}
Since $\tau_{i,ind}=dL_{i,ind}/dt$, we obtain
\begin{equation}\label{eq4}
\tau_x = (I_z \omega_z)\Omega_y + I_x \dot{\Omega}_x.
\end{equation}
Applying the same argument, for the y-component we obtain
\begin{equation}\label{eq5}
\tau_y = -(I_z \omega_z)\Omega_x + I_y \dot{\Omega}_y.
\end{equation}
The equations of change of angular momentum can be written under the compact form:
\begin{equation}\label{eq5a}
\tau_i = \sum_{jk} \left[  \dot{\alpha} \frac{\partial}{\partial \alpha_k} + \dot{\omega}_k \frac{\partial}{\partial \omega_k} \right]I_{ij} \omega_j
\end{equation}
where $\mathbf{\alpha}=(\alpha_1,\alpha_2,\alpha_3)$ represents Eulerian angles. By using a statistical description of a set of rotating bodies~\cite{Curtiss 1956} (e.g., tornadoes, microscopic particles in a plasma showing diamagnetic behavior) introducing a a distribution function in the generalized space $f(\mathbf{r},\mathbf{v},\pmb{\alpha},\pmb{\omega},t)$, such that
\begin{equation}\label{eq5b}
f(\mathbf{r},\mathbf{v},\pmb{\alpha},\pmb{\omega},t) d \mathbf{r} d \mathbf{v} d \pmb{\alpha} d \pmb{\omega}
\end{equation}
represents the number of rotating bodies with vector position $\mathbf{r}$ between $\mathbf{r}$ and $\mathbf{r}+d\mathbf{r}$, with velocity $\mathbf{v}$ between $\mathbf{v}$ and $\mathbf{v}+d\mathbf{v}$, with $\pmb{\alpha}$ between $\pmb{\alpha} +d\pmb{\alpha}$, and with $\pmb{\omega}$ between $\pmb{\omega}+d\pmb{\omega}$. Here, the symbol $\pmb{\alpha}$ denotes simply a functional dependence on Eulerian angles $\alpha, \beta$ (see Fig.~\ref{fig1}). The distribution function is normalized:
\begin{equation}\label{eq5c}
\int f d\mathbf{v} d\pmb{\alpha} d\pmb{\omega}=n,
\end{equation}
with $n$ denoting the number density of rotating bodies. For convenience we may define an average over $\mathbf{v}$, $\pmb{\alpha}$, and $\pmb{\omega}$ and denote it by
\begin{equation}\label{eq5d}
\widetilde{A}=\frac{1}{n} \int \int \int A(\mathbf{v},\pmb{\alpha},\pmb{\omega}) f(\mathbf{r},\mathbf{v},\pmb{\alpha},\mathbf{\omega},t) d \mathbf{v} d\pmb{\alpha} d \pmb{\omega}.
\end{equation}

\begin{figure}
  \includegraphics[width=4.0 in]{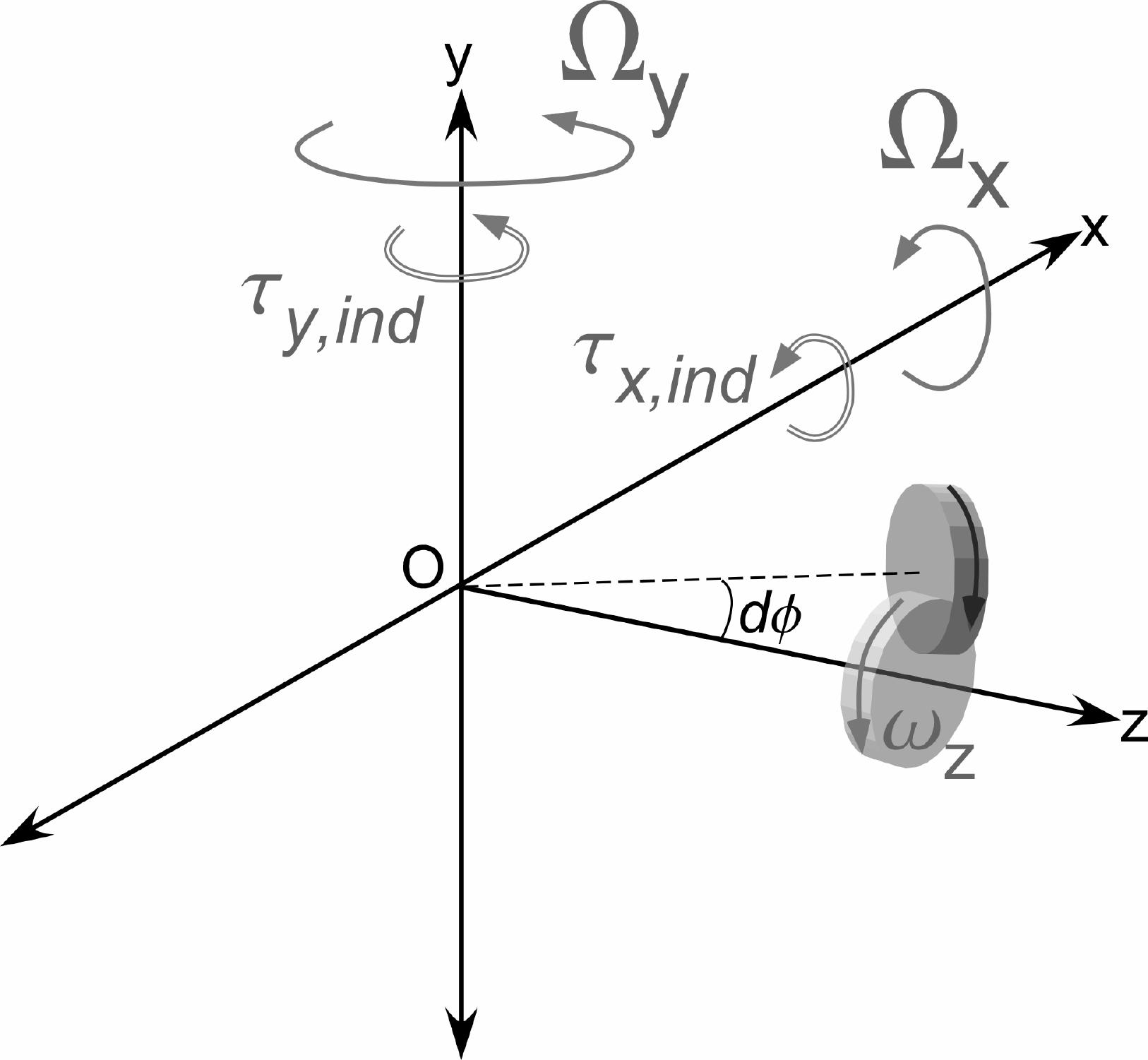}\\
  \caption{Gyroscopic motion of a wheel rotating with angular velocity $\omega_z$ around its main axis $O_z$, and angular velocity $\Omega_y$. Notice that the induced torque is transverse to the imposed displacement.}\label{Fig1}
\end{figure}

The above statistical treatment (see Ref.~\cite{Curtiss 1956} and Eq.2-51 therein) leads us to the equation of conservation of angular momentum, which can be written under the form:
\begin{equation}\label{eq5e}
\frac{\partial }{\partial t} \left( n \widetilde{\mathbf{M}} \right) + \nabla \cdot (n \mathbf{v}_0 \widetilde{\mathbf{M}}) = - \nabla \cdot \overleftrightarrow{\Phi} - \nabla \cdot \overleftrightarrow{\Phi_0} + n\mathbf{\tau},
\end{equation}
where we have introduced the following dyadics
\begin{equation}\label{eq5f}
\begin{array}{c}
\overleftrightarrow{\Phi}  = n\widetilde{\mathbf{V}(\mathbf{I} \cdot \pmb{\Omega})}; \\
\overleftrightarrow{\Phi_0}= n\widetilde{\mathbf{V}(\mathbf{I} \cdot \pmb{\omega}_0)},
\end{array}
\end{equation}
which is the tensor representing the flux of angular momentum $\pmb{\Omega}=\pmb{\omega}-\pmb{\omega_0}$, with $\pmb{\omega}_0$ denoting the average angular velocity of the set of rotating bodies (the same tensor applies to $\pmb{\omega_0}$); $\mathbf{v}_0$ is the macroscopic stream velocity of all the rotating bodies, and $\mathbf{V}=\mathbf{v}-\mathbf{v}_0$ is the velocity of a rotating body relative to the stream. If we assume that the set of rotating bodies is at rest, Eq.~\ref{eq5d} simplifies to
\begin{equation}\label{eq5g}
\frac{\partial }{\partial t} \left( n \widetilde{\mathbf{M}} \right) + \nabla \cdot (n \widetilde{\mathbf{v}(\mathbf{I} \cdot \mathbf{\omega}}) = n\mathbf{\tau}.
\end{equation}
Integration over a volume yields
\begin{equation}\label{eq5h}
\int \int \int_V \frac{\partial }{\partial t} \left( n \widetilde{\mathbf{M}} \right)d\mathbf{r} + \int\int\int_V \nabla \cdot (n \widetilde{\mathbf{v}(\mathbf{I} \cdot \mathbf{\omega}}) d\mathbf{r} = \int\int\int_V n\mathbf{\tau} d\mathbf{r}.
\end{equation}
The second term on the left hand side can be transformed into a surface integral; therefore, Eq.~\ref{eq5h} can be written as
\begin{equation}\label{eq5i}
\frac{\partial \widetilde{\mathbf{M}}}{\partial t} + \oint_S \overleftrightarrow{\Phi} \cdot \mathbf{n}^* d S=\mathbf{\tau}.
\end{equation}
If we substitute in the surface integral $n\mathbf{V}$ by an operator:
\begin{equation}\label{eq5k}
n \mathbf{V} \cdot \equiv \sigma \frac{\partial}{\partial t} \cdot,
\end{equation}
where $\sigma =N/S$, and $N$ is the total number of rotating bodies inside the volume limited by $S$, then Eq.~\ref{eq5i} reduces to the expression:
\begin{equation}\label{eq5l}
\frac{\partial \widetilde{\mathbf{M}}}{\partial t} + \sigma \frac{\partial}{\partial t} \oint_S (\mathbf{L} \cdot \mathbf{n}^*) d S=\mathbf{\tau}.
\end{equation}
Here, the second integral was converted to an integral over an inner product of vectors in three-dimensional space, similar to Faraday's law of induction. We have also used the geometric property $dS_j=n_jdS \leftrightarrow d\mathbf{S}=\mathbf{n}^{*}dS$.

\begin{figure}
  \includegraphics[width=4.0 in]{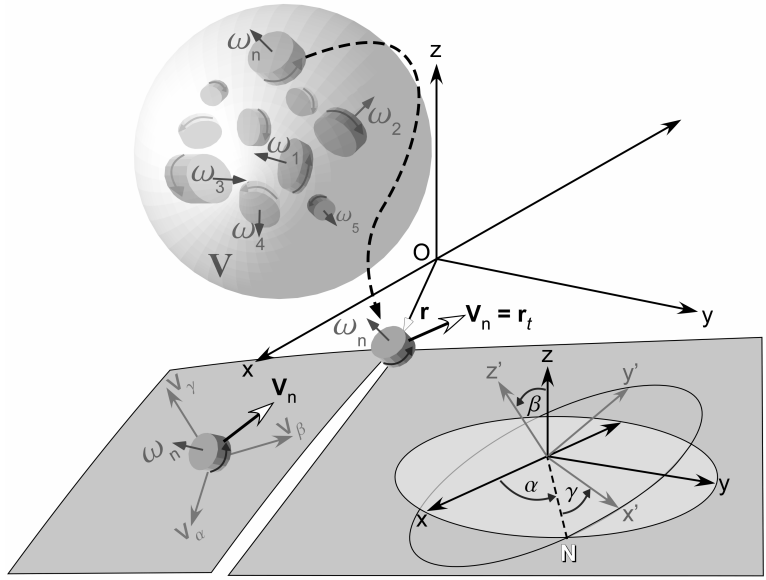}\\
  \caption{A volume $\mathbf{V}$ containing a set of rotating bodies, each with vorticity $\omega$. In particular, the figure represent the velocity components and Euler's angles for one of bodies.}\label{fig1}
\end{figure}

\begin{figure}
  \includegraphics[width=4.0 in]{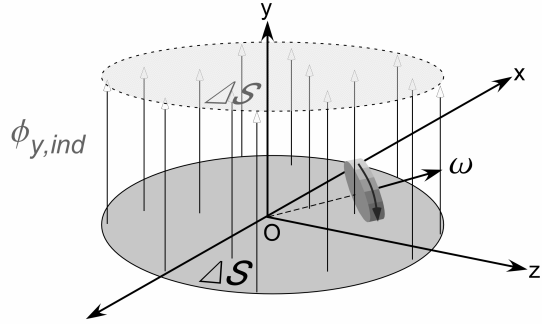}\\
  \caption{Axial flux of angular momentum $\phi_{y,ind}$ through an area $\Delta S$.}\label{Fig4}
\end{figure}

In Eqs.~\ref{eq4}-~\ref{eq5}, $I_x$ and $I_y$ are the inertial momentum of the rotor (wheel) relative, resp., to the $O_x$ and $O_y$ axis.
It is clear that the last term can be represented by a time rate of change of a flux of angular momentum (per unit of area):
\begin{equation}\label{eq6}
\Phi_y = \frac{1}{S} \oint (\mathbf{L}_{ind} \cdot d \mathbf{S}).
\end{equation}
We may use here the concept of the angular momentum field lines threading an hypothetical surface whose boundary is an open surface $S(t)=\oint dS=2 \pi (1-\cos \alpha)$ (in fact, the area of the spherial cap), analogously to the procedure used in Faraday's law of induction. The ``angular momentum field" is supposed to have the origin at the extremity coincident with the gyroscope main axis inclined at an angle $\alpha$ relative to the vertical (if subject to the gravity field). This procedure is offered to manipulate mathematically Eq.~\ref{eq6}, see also Fig.~\ref{Fig4}.

However, the angular momentum flux threading the surface $d \mathbf{S}$ is homogeneous, and Eq.~\ref{eq6} gives simply
\begin{equation}\label{eq7}
\Phi_i = L_{ind,i}=I_i \Omega_i,
\end{equation}
for $i=1,2,3$. This ``phase" is mass dependent in a way which seems consistent with the quantum mechanics prediction that all phase-dependent phenomena (even in the presence of a gravitational field) depend on the mass through de Broglie's wavelength~\cite{Overhauser}, as we will see later.
We can rewrite Eq.~\ref{eq3} under the general form:
\begin{equation}\label{eq8}
\tau_i = \epsilon_{ijk} \Omega_j A_k + \partial_t \Phi_i.
\end{equation}
where the inferior Latin index $i=1,2,3$ refers to the Cartesian components of the physical quantities along the axes of the inertial system $1,2,3$; $\epsilon_{ijk}$ denotes the completely anti-symmetric unit tensor of Levi-Civita; $A_k=\omega_k I_k$; and the unit vector $\overrightarrow{n}$ points along the $O_x$ or $O_y$-axis.

From this analysis we propose a new law for the direction of precession: if a moving body is acted upon either gradually or with a sudden jerk, the resultant angular momentum flux through a closed linear momentum isosurface induces an angular momentum to compensate such a change, that is, there is a tendency to retain a constant value of flux as described in Eq.~\ref{eq6}. the last term of Eq.~\ref{eq7} is related to the Faraday's magnetic induction law by considering the analogy between the magnetic field and vorticity~\cite{MPinheiro}. quite interestingly, the above construction also implies a relationship similar to the Lenz law.

From this relationship, one can predict the direction of precession for an axisymmetric gyroscope with high angular velocity: if we assume that $\pmb{\omega}$ is pointing up, and we increase the value of a suspended mass, then
the precessional velocity turning to the right will increase, since the vector weight points down and would reduce the value of the angular momentum. Therefore, there must be a compensating angular momentum through precession in the same direction as $\pmb{\omega}$, in order to maintain constant flux; on the contrary, if $\pmb{\omega}$ points down, the precession motion would reverse direction. We encourage the reader to test with a simple ring gyroscope. This phenomena implies a deep connection between inertia, gravity and the physical vacuum. This interpretation bears some resemblance to a fluidic spacetime interpretation proposed by Meholic~\cite{Greg}, interpreting inertia as caused by the time lag effects of the spacetime medium reacting to a moving gravitational source).


\begin{figure}
  \includegraphics[width=3.0 in]{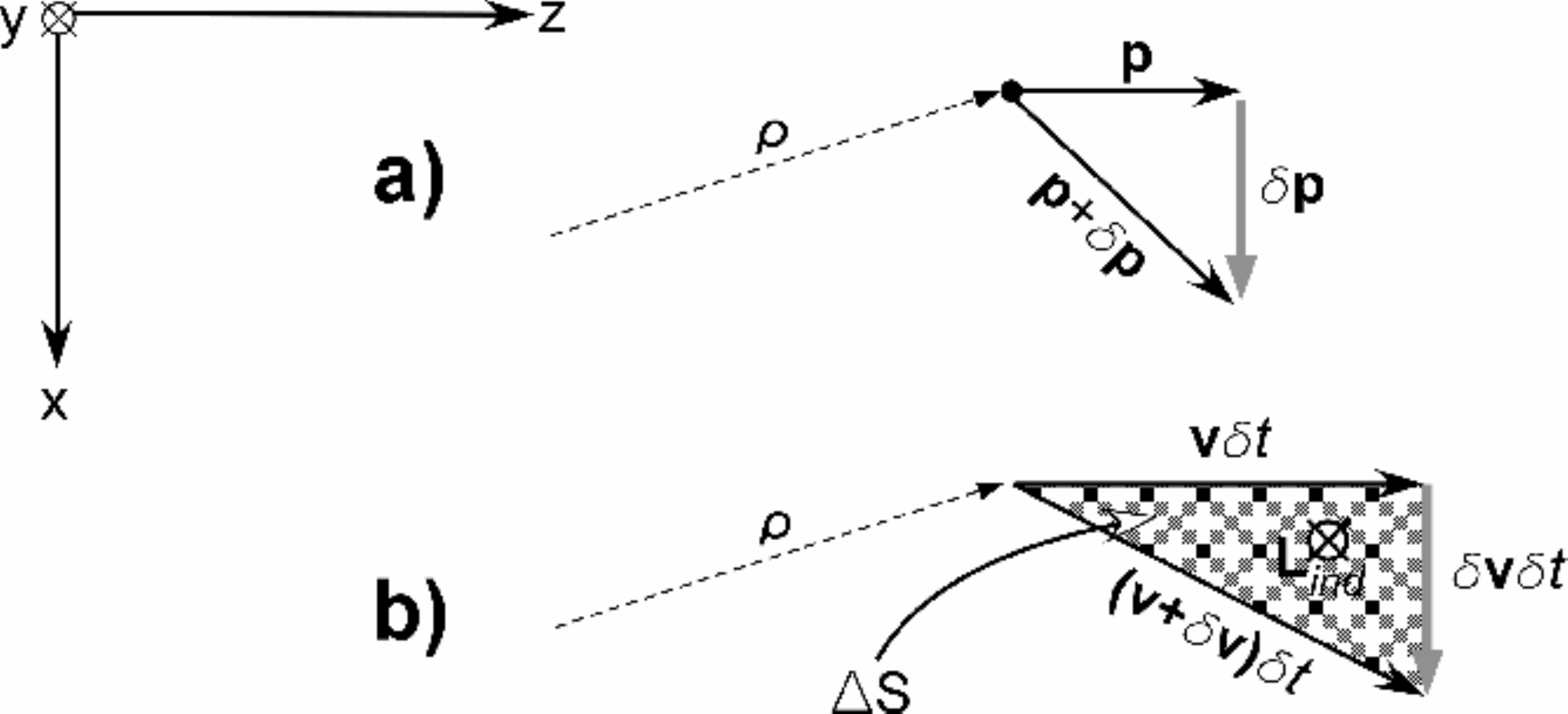}\\
  \caption{(a) - A particle with initial momentum at position $\rho$ experiences an imposed acceleration; (b) - the induction law implies an induced angular momentum $\mathbf{L}_{ind}$ "threading" the region $\Delta S$.}\label{Fig1}
\end{figure}

A similar analysis can be done relative to the fundamental equation of dynamics. For this purpose let us examine Fig.~\ref{Fig1}-(a). Consider a particle of mass $m$ with linear momentum $\mathbf{p}(t)$ moving along the $O_x$ axis. Let us suppose that it experiences a sudden change of linear momentum due to action of some agent, $\delta \mathbf{p}$. Consequently, at time $t+ \Delta t$ obtains a new linear momentum $\mathbf{p}(t+\Delta t)$. There is always a given point in space distant of $\rho$ relative to which the particle has angular momentum. The important point to note is that this sudden change of motion (i.e., trajectory) is in fact equivalent to a succession of infinitesimal rotational displacements and, therefore, related to a sudden appearance of angular momentum $\delta \mathbf{L}=\rho \delta p \mathbf{u}_y$ (where $\mathbf{u}_y$ is the unit vector), about an instantaneous axis the position of which passes through the point of intersection of the linear momentum at previous times $t$ and $t+\tau$. Then, it can be shown (see Fig.~\ref{Fig1}-(a)) that
\begin{equation}\label{eq9}
\delta \mathbf{p} = \frac{1}{\rho} (\delta \mathbf{L} \cdot \mathbf{u}_y),
\end{equation}
or
\begin{equation}\label{eq10}
 \partial_t \delta \mathbf{p} = \frac{1}{\rho} ( \partial_t \delta \mathbf{L} \cdot \mathbf{u}_y),
\end{equation}
for a given $\rho$. One then easily finds that, besides an eventual external force $\mathbf{F}^{ext}$, {\it there suddenly} appears a new force term:
\begin{equation}\label{eq11}
m \mathbf{a} = \mathbf{F}^{ext} + \partial_t \delta \mathbf{p} .
\end{equation}
The angular momentum is (through the agency of $\pmb{\omega}$) the analogue of the B-field (see., e.g., Ref.~\cite{MPinheiro}), hence it is natural to associate with $\mathbf{L}$ an induction law. This is manifested through a resistance to the sudden increase of angular momentum flux threading an elementary area $dS$ (as we have discussed previously). This is quantified through the relationship:
\begin{equation}\label{eq12}
m \mathbf{a} = \mathbf{F}^{ext} + \partial_t \left( \delta \mathbf{p} \right)_{ind},
\end{equation}
with $\partial_t \left(\delta \mathbf{p} \right)_{ind} = \partial_t \left(\delta \mathbf{p} \right)$, representing the inertial resistance with which a single mass particle opposes the imposed acceleration. Note that, when performing an infinitesimal change of time, then we have
\begin{equation}\label{eq14}
\delta \mathbf{L} = \mathbf{L}(t + \tau ) - \mathbf{L}(t)\approx \tau \frac{\partial \mathbf{L}}{\partial t},
\end{equation}
which gives
\begin{equation}\label{eq15}
\frac{\partial \delta \mathbf{L}}{\partial t} \approx \tau \frac{\partial^2 \mathbf{L}}{\partial t^2},
\end{equation}
for a given $\tau$. Assuming that the ``surge" directed along the $O_x$ axis induces angular momentum along the $O_y$ axis, we obtain
\begin{equation}\label{eq16}
\frac{\partial \delta \mathbf{L}}{\partial t} \approx \tau I_y \ddot{\Omega}_y \mathbf{u}_y.
\end{equation}
Here, $I_y=m\rho^2$. Hence, since $\Omega_y=v_x/\rho$
\begin{equation}\label{eq17}
\frac{1}{\rho} \left( \frac{\partial \delta \mathbf{L}}{\partial t} \cdot \mathbf{u}_y  \right)=m \tau \dot{a}_x.
\end{equation}
Therefore, we deduce immediately that the following equation holds:
\begin{equation}\label{eq18}
m a_x=F^{ext}_x + m \tau \dot{a}_x.
\end{equation}
We can write it in a general form:
\begin{equation}\label{eq19}
m a_i = F_i^{ext} + m \tau \dot{a_i},
\end{equation}
with $i=1,2,3$. This is an equation similar to the Abraham's equation that explains (classically) the radiation problem. A relationship can be inferred for $\tau$. The {\it critical action time} $\tau \equiv \aleph/m$ is dictated by the kind of physical fields acting over the body. For example, a simple argument based on the charged particle canonical momentum $\delta \overrightarrow{p} = m \delta \overrightarrow{v} + q \delta \overrightarrow{A}$ leads us to a rough estimation of $\tau$ when considering radiation problems. The electromagnetic vector potential produced by a charge $q$ in motion with speed $v$ at a distance $r$ is $A=qv/4 \pi \epsilon_0 c^2 r$. We can expect that a sudden change of velocity leads to $r \approx e^2/mc^2$, where $e^2=q^2/4 \pi \epsilon_0$. The critical action time (or transient time) $\tau$ is such as
\begin{equation}\label{eq20}
\tau =\frac{r}{c} \approx \frac{e^2}{mc^3},
\end{equation}
which is of the order of value $\tau=2 e^2/3mc^3$ given in the framework of the Abraham model. And $\tau$ measures the significance of the electromagnetic field relatively to the inertial mass. Notice that it is not necessary that the particle (or body) rotates in order to obtain angular momentum; it is enough that there exists a variation of $\overrightarrow{L}$ relative to a given point (or axis).

Quite interestingly, the inductive term can be identified with the Berry's geometric phase~\cite{MBerry} acquired over the course of a cycle, when the system is subjected to cyclic adiabatic processes. In the case of the Aharonov-Bohm effect, the adiabatic parameter is the magnetic field inside the solenoid, and the rotation around a given axis corresponds to a closed loop. Hence in the study of the generalized top, the inductive term generates a torque, which is a similar process to the observation of currents induced due to Faraday's law. We can show next that the Aharonov-Casher flux effect has the same origin as the last term of Eq.~\ref{eq12}.

For this purpose, let us consider a mesoscopic ring of radius $R$ the AC flux is given by~\cite{Balatsky}:
\begin{equation}\label{eq21}
\frac{\Phi_{AC}}{\Phi_o} = g \mu_B \frac{RE}{c \hbar},
\end{equation}
where $\Phi_o=h/e$, $g$ is the g-factor, $\mu_B$ is the Bohr magneton, $E$ is the electric field actuating upon the particle in the ring undergoing rotational motion relative to a $z$ axis of revolution draw along its axis of revolution. Inserting the Bohr magneton $\mu_B=eL/2m$ into Eq.~\ref{eq21} we obtain
\begin{equation}\label{eq22}
\frac{\Phi_{AC}}{2 \pi} = \frac{g}{2}\frac{eREL}{emc}.
\end{equation}
However, $L=I_z\Omega_z$, with $I_z=mR^2$ since the particle of mass $m$ orbits along the ring of radius $R$. In addition, note that we can scale $eER$ with the rest energy of the particle $mc^2$~\cite{Balatsky}. This gives
\begin{equation}\label{eq23}
\tau \Phi'_{AC} = \frac{g}{2} m R^2 \Omega_z = \frac{g}{2} I_z \Omega_z.
\end{equation}
Here, we have made the appropriate change of variable $\Phi'_{AC}=c^2 \Phi_{AC}/2 \pi me$, in order to get back our ``phase" variable in angular momentum units. Therefore, this expression is identical to the inductive term:
\begin{equation}\label{eq24}
\tau \frac{d \Phi'_{AC}}{dt} = \frac{g}{2} I_z \dot{\Omega}_z.
\end{equation}

Since space does not offer any resistance to bodies with uniform motion, we can deem the inductive term of Eq.~\ref{eq1} as the cause of inertia, implying that space itself resists a change of angular momentum. This backreaction of space on the electrons inside the mesoscopic ring (described previously) is similar to what happens to the particles comprising the gyroscope.

We emphasize that this relationship is not describing the action of the electromagnetic field, as in the case of Faraday's law of induction, but is applied solely to mechanical motion through the proposed mechanism.

This implication goes beyond abstract mathematics and probes the fundamental nature of the vacuum: it seems to possess a physical structure, with a back-reaction that acts only on quasi-2D cross-sections.

Although Einstein in the frame of the special theory of relativity disregarded the notion of absolute space, he returned to this concept later when formulating his general theory of relativity, by the spacetime metric $g_{\mu,\nu}$, showing that spacetime has a geometrical structure, and claiming that empty space should be called the ``the new aether of the general relativity"~\cite{Overduin} (see also Ref.~\cite{Muller}).

Much earlier, Newton showed with his bucket experiment that when water climbs the walls of the bucket, it is an indication that water ``knows" it is spinning with respect to absolute space (Newton employs this term for the medium which fills space). Newton gave a procedure to test if one is moving uniformly with respect to absolute space: by checking if Newton's law of motion was satisfied.
Of course, Mach attempted to describe the world exclusively in terms defined relative to material bodies and explained Newton's bucket experiment differently, arguing that the water rises up the walls of the bucket due to influence exerted by the rest of the matter of the universe. The present discussion tends to show, however, that inertia is due to local interactions.

Besides possible propulsion issues that can be envisaged from the explanation given here, this framework could be relevant for electromagnetic and gravitational wave generation~\cite{Fontana}.

\section{Conclusion}

When a system is subject to a sudden change of velocity or direction of motion, there appears an induction force that opposes such a change. This induction force is related to the inertia of bodies. The previous discussion implies that a nonuniform motion of a body that is under observation from a frame of reference (inertial or noninertial), undergoes real physical changes due to the intrinsic properties of the physical vacuum, its polarizability and back-reaction inductive term. The inductive term can be identified with the Berry's geometric phase acquired over the course of a cycle, when the system is subjected to cyclic adiabatic processes. As for the electromagnetic field, the inductive effect is local and referred to absolute space, apparently countering Mach's viewpoint~\cite{Pinheiro}. This work reveals a new concept of unity among the fundamental physics that govern classical mechanics and electrodynamics, from macroscopic to microscopic scales.

\begin{acknowledgments}
One of us (MJP) gratefully acknowledge partial financial support by the International Space Science Institute (ISSI) as visiting scientist and express special thanks to Prof. Roger Maurice-Bonnet.
\end{acknowledgments}

\bibliographystyle{amsplain}

\end{document}